\documentclass{article}

\usepackage{arxiv}

\usepackage[utf8]{inputenc} % allow utf-8 input
\usepackage[T1]{fontenc}    % use 8-bit T1 fonts
\usepackage{hyperref}       % hyperlinks
\usepackage{url}            % simple URL typesetting
\usepackage{booktabs}       % professional-quality tables
\usepackage{amsfonts}       % blackboard math symbols
\usepackage{nicefrac}       % compact symbols for 1/2, etc.
\usepackage{microtype}      % microtypography
\usepackage{lipsum}
\usepackage{graphicx}
\usepackage{caption}
\usepackage{bm}

\title{\huge The Koha Code \\ \LARGE A Biological Theory of Memory}

\author{
  Lum Ramabaja
}

\begin{document}
\maketitle

\begin{abstract}

This work introduces the Koha model, a new theory that aims to explain two unresolved phenomena within biological neural networks: 
How information is processed and stored within neural circuits, and how neurons learn to become pattern detectors. In the Koha model, the dendritic spines of a neuron serve as computational units that scan for precise spike patterns in their synaptic inputs. The model proposes the existence of a temporal code within each dendritic spine, which is used for the dampening or amplification of signals, depending on the temporal information of incoming spike trains. Compelling evidence is provided and a concrete process is described for how signal filtration occurs within spine necks. A competitive learning algorithm is then proposed that describes how neurons use their internal temporal codes to become pattern detectors.

\end{abstract}

% keywords can be removed

\section{Introduction}
\label{s: intro}

\subsection{Form follows function}
\label{ss: form follows function}
The ability to form and store memories, recall past events, make inferences and deductions, are all examples of "emergent algorithms" that arise from neural circuits. 
Emergence occurs when a system exhibits characteristics that its constituents do not possess but emerge because of rules, or interactions between its parts. Parts of an emergent system often adhere to very basic principles. When those parts interact with one another, complexity can arise. In the case of the brain, neural circuits serve as the physical substrate for emergent algorithms, which arise as a result of neuronal interactions. Any modification to a neural circuit or structure may thus affect an existing algorithm or, in certain instances, result in the creation of a new algorithm. If a useful algorithm forms as a result of a mutation in one of its underlying neural circuits, the organism with the altered algorithm will have a competitive advantage over other organisms. Over time, evolutionary forces will shape the underlying circuits in a way that makes the emergent algorithm more efficient. This means that by examining the optimized "hardware" of organisms, it might be possible to deduce some of the existing emergent algorithms in nature. If a certain pattern repeats across many species, we can be very certain that the repeating pattern plays a significant role. By looking at the optimised morphology, structures, and processes of various neurons, we can start to guess their purpose in the context of the mind. After all, in biology, form follows function.

\subsection{Patterns in the substrate}
\label{ss: patterns in the substrate}
One of the most notable exitatory neurons is the pyramidal neuron. It is the most common type of excitatory neuron in the mammalian cortical structures, as well as one among the largest neurons in the brain. Each mammal, as well as birds, fish, and reptiles, has pyramidal neurons. Apart from its wide occurrence in nature, the pyramidal neuron exhibits also many noteworthy physical features, particularly when contrasted to inhibitory interneurons. The following is a list of key observations that are relevant for the Koha model presented in section \ref{s:the_hypothesis}:

\begin{itemize}
    \item Studies in the layer 2/3 of the rat neocortex have shown that pyramidal neurons posses about 10 times more local connections with fast-spiking inhibitory neurons, than with other pyramidal neurons, with most of the connections being reciprocal \cite{holmgren_pyramidal_2003}.
    
    \item Pyramidal neurons have two kinds of dendrites. \textit{Basal dendrites}, which come out of the soma and branch extensively. And a single, long, dendrite that emerges from its soma, also known as the \textit{apical dendrite}. It can extend for several hundred microns before branching into many smaller dendrites (see figure \ref{fig:pyramidal_neuron1}) also known as the \textit{apical tuft}. Basal dendrites tend to receive their input from nearby neurons, whereas the apical dendrite receives its inputs from more distant regions \cite{henze_dendritic_1996}. While pyramidal neurons have an apical dendrite, inhibitory interneurons do not.
    \begin{minipage}{\linewidth}
    \centering
    \captionsetup{type=figure}
    \includegraphics[width=0.8\linewidth]{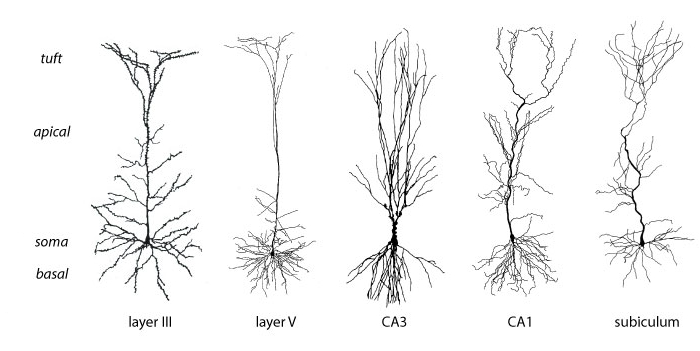}
    \caption{The variable structure of pyramidal neurons across different parts of the brain. Even though the pyramidal neuron differs slightly across layers, it is characterized by its stereotypical morphology - the presence of an apical dendrite and the presence of basal dendrites. Adapted from Spruston, 2008 \cite{spruston_pyramidal_2008}}.
    \label{fig:pyramidal_neuron1}
    \end{minipage}

    \item The apical dendrite and basal dendrites of pyramidal neurons are completely covered with tiny protrusions structures known as \textit{dendritic spines} (see figure \ref{fig:spine1}). It is believed that dendritic spines serve as memory storage components for neurons \cite{yuste_dendritic_2010}. In contrast to excitatory neurons, most interneurons do not posses, or posses very few dendritic spines, i.e. most interneurons are essentially spineless \cite{yuste_dendritic_2010}.
    \begin{minipage}{\linewidth}
    \centering
    \captionsetup{type=figure}
    \includegraphics[width=0.6\linewidth]{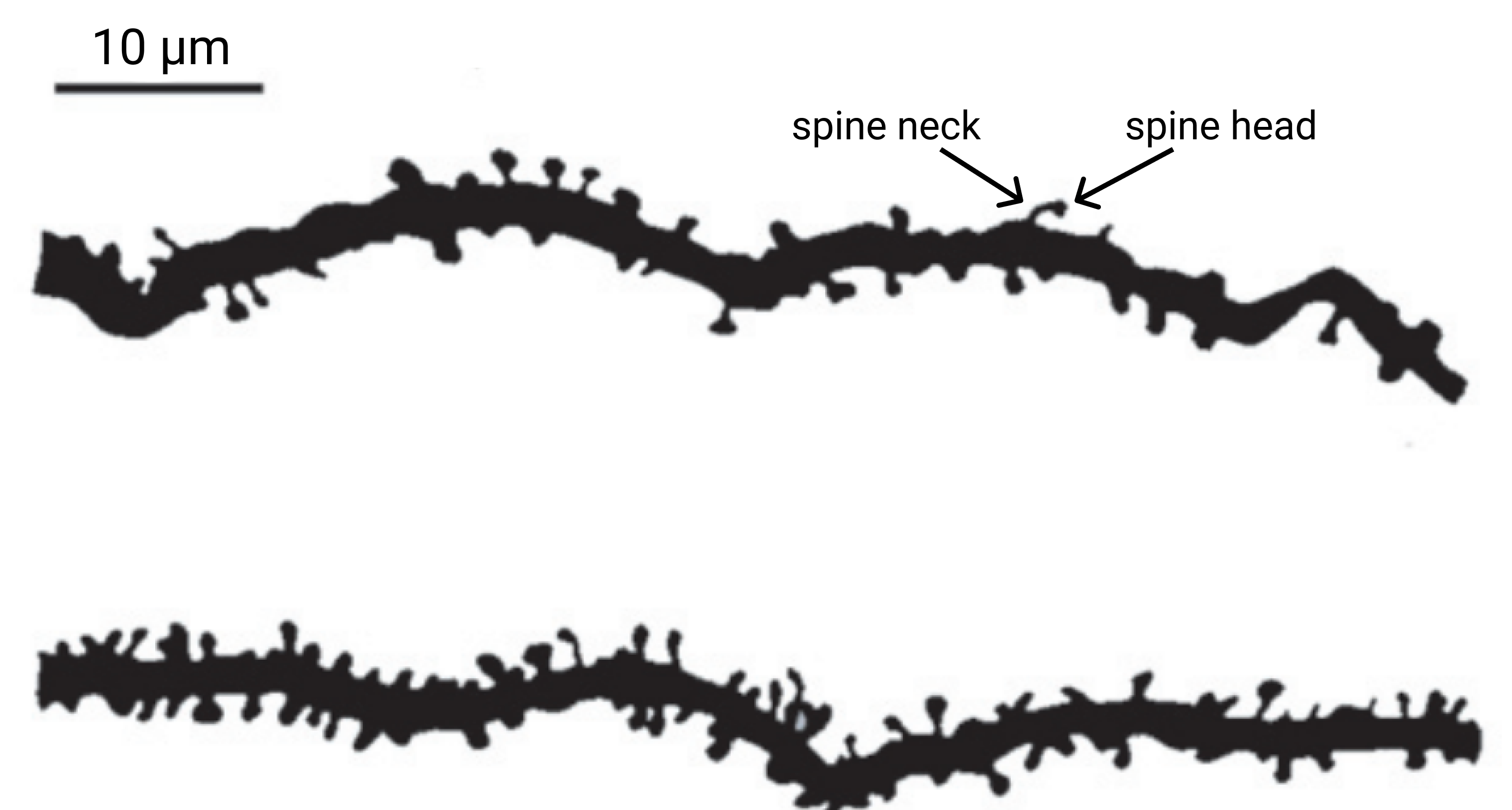}
    \caption{Two dendrites covered with dendritic spines. Adapted from Spruston, 2008 \cite{spruston_pyramidal_2008}}.
    \label{fig:spine1}
    \end{minipage}

    \item There are various stages of dendritic spines, the exact borders of which are a bit blurry. Traditionally spines are classified into thin, stubby, and mushroom spines \cite{peters_small_1970} (see figure \ref{fig:spine2}). Thin spines have a small head and a long neck that connects to the parental dendrite. Stubby spines have no neck at all and are more expressed in infants \cite{harris_three-dimensional_1992}. Whereas mushroom spines have a wide neck with a large head and are more expressed in adult brains \cite{harris_three-dimensional_1992,berry_spine_2017}. Because of the properties each spine category has, it is believed that thin spines may play a role in "learning", whereas mushroom spines in long-term memory \cite{bourne_thin_2007}. Thin spines can dynamically appear and disappear throughout life, while most mushroom spines remain stable for a lifetime \cite{bourne_thin_2007, holtmaat_transient_2005}. The stability of the mushroom spine and its high expression in adult brains, make it an ideal candidate as a long term "memory unit".
    \begin{minipage}{\linewidth}
    \centering
    \captionsetup{type=figure}
    \includegraphics[width=0.4\linewidth]{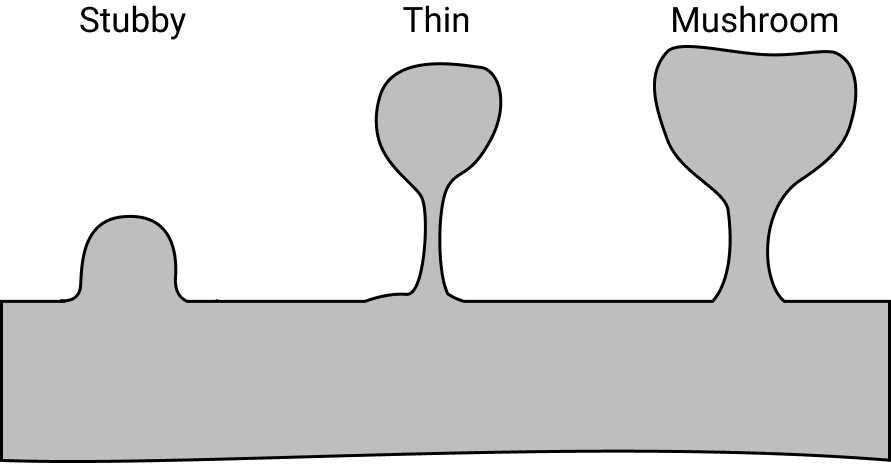}
    \caption{The three traditional types of dendritic spines.}
    \label{fig:spine2}
    \end{minipage}

    \item We now know that dendritic spines receive most of the excitatory inputs in neurons, and that practically every dendritic spine has an excitatory synapse on its head \cite{gray_electron_1959, arellano_non-synaptic_2007, colonnier_synaptic_1968}. In other words, excitatory axons prefer for some reason to terminate particularly on spines, they almost never connect directly to dendritic shafts (the body of the dendrite) \cite{colonnier_synaptic_1968}. We can see the opposite behavior in inhibitory axons. Inhibitory axons almost always connect directly to dendritic shafts, rather than on spines \cite{somogyi_salient_1998}. The consistency of this pattern suggests a functional value.
    
    \item Imaging experiments have demonstrated that dendritic spines behave indeed like electrical and biochemical compartments \cite{volfovsky_geometry_1999, wickens_electrically_1988, lisman_mechanism_1989, koch_function_1993, holmes_is_1990, volfovsky_geometry_1999}. The narrow neck of the spine can create an isolated compartment in which the biochemical signals of the spine head do not spread along the parent dendrite. Changes in the length and narrowness of the spine neck, has a direct effect on the compartmentalization of the spine \cite{basu_role_2018}.
    
    \item Experiments also show that the most significant filtering of local potentials in a neuron does not occur along the dendrite, but instead at the spine neck \cite{araya_spine_2006}. The spine neck is somehow able to modulate the amplitude of incoming signals. Further studies have also shown that spines, additionally to filtering out local potentials, are also able to amplify them \cite{araya_sodium_2007}.
    
    \item Evidence is now overwhelming that dendritic spines can change their physical structure in seconds, following an appropriate stimulus \cite{yuste_mechanisms_1999, yuste_dendritic_1995, fischer_rapid_1998, dunaevsky_developmental_1999, korkotian_bidirectional_1999}. Spine necks have an actin cytoskeleton (see figure \ref{fig:actin_spine_neck}) that can dynamically change shape depending on its interactions with several actin regulators. This change in geometry has a direct affect on local voltage amplification and biochemical compartmentalization \cite{noguchi_spine-neck_2005}. The strength of incoming signals can therefore be dampened, or amplified, depending on the length and width that the spine neck takes \cite{araya_activity-dependent_2014}.
    \begin{minipage}{\linewidth}
    \centering
    \captionsetup{type=figure}
    \includegraphics[width=0.4\linewidth]{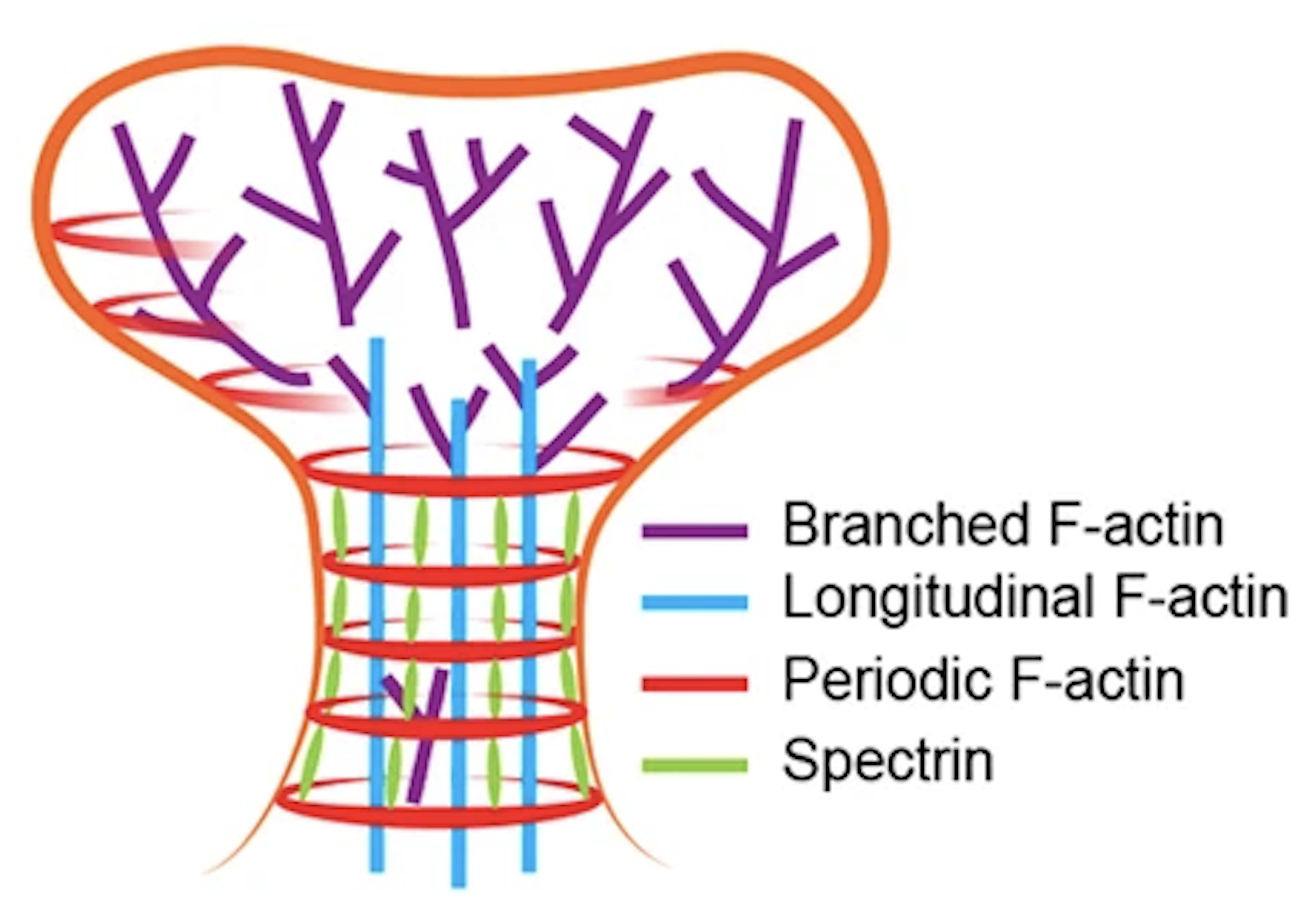}
    \caption{A simplified model of the organization of actin in dendritic spines. Reprinted from \cite{bar_periodic_2016}}.
    \label{fig:actin_spine_neck}
    \end{minipage}

    \item Some dendritic spines also posses an enigmatic organelle known as the spine apparatus, a specialized formation of stacked endoplasmic reticulum that occupies a large portion of the spine neck. Not all spines have a spine apparatus, but almost all mushroom spines have one \cite{spacek_three-dimensional_1997}. We now know that the role of the spine apparatus is closely linked with local calcium trafficking \cite{fifkova_possible_1985, korkotian_fast_1998, sharp_differential_1993}. Due to the large surface area of the organelle, it is believed that the spine apparatus may act as a calcium buffer \cite{fifkova_possible_1985}. Studies now show that the spine apparatus is also able to quickly release $Ca^{2+}$ \cite{verkhratsky_physiology_2005, holbro_differential_2009}. The spine apparatus is thus acting as an intracellular calcium store \cite{fifkova_possible_1985}, serving as both calcium source and calcium sink, depending on internal feedback. 
    
    \item The spine apparatus also appears to be closely associated with synaptopodin \cite{deller_actin-associated_2000}, which is an Actin-associated Protein. In an experiment, synaptopodin-deficient mice were associated with a loss of the spine apparatus and showed a reduction in hippocampal long-term potentiation, suggesting that the spine apparatus may play an important role in synaptic plasticity \cite{deller_synaptopodin-deficient_2003}. Synaptopodin's preferential location in spines, and its close association with the spine apparatus and the actin cytoskeleton, suggests that synaptopodin is somehow involved in the rapid plasticity of the spine's actin cytoskeleton. In fact, one study suggested that synaptopodin may act as an actin-bundling molecule in the spine neck, and that it may influence calcium release from the spine apparatus \cite{deller_potential_2000}.  
    
    \item An even more engimatic organelle is the cisternal organelle, found at the axon initial segment of pyramidal neurons. The cisternal organelle is also a formation of stacked endoplasmic reticulum and shares many structural similarities with the spine apparatus. Very little is known about this organelle. There is now evidence that it may serve as an intracellular calcium store \cite{benedeczky_cisternal_1994}. The cysternal organelle, just as the spine apparatus, also appears to be closely associated with synaptopodin. Experiments showed that synaptopodin-deficient mice were also associated with a loss of the cisternal organelle at the axon initial segment \cite{bas_orth_loss_2007}.

\end{itemize}

\subsection{Receptive fields and invariances}
\label{ss: receptive fields and invariances}
The portion of sensory space that can cause a neuron to fire when stimulated is known as the neuron's \textit{receptive field}. There are visual receptive fields, auditory receptive fields, olfactory receptive fields, and somatosensory receptive fields. In the 1950s and 1960s, Hubel and Wiesel demonstrated that different visual stimuli have different effects on the firing pattern of a neuron \cite{hubel_receptive_1959, hubel_receptive_1968}. Hubel and Wiesel categorized the visual neurons in their experiments into two groups: \textbf{simple cells} and \textbf{complex cells}. 

\begin{figure}[h]
\centering\includegraphics[width=0.4\linewidth]{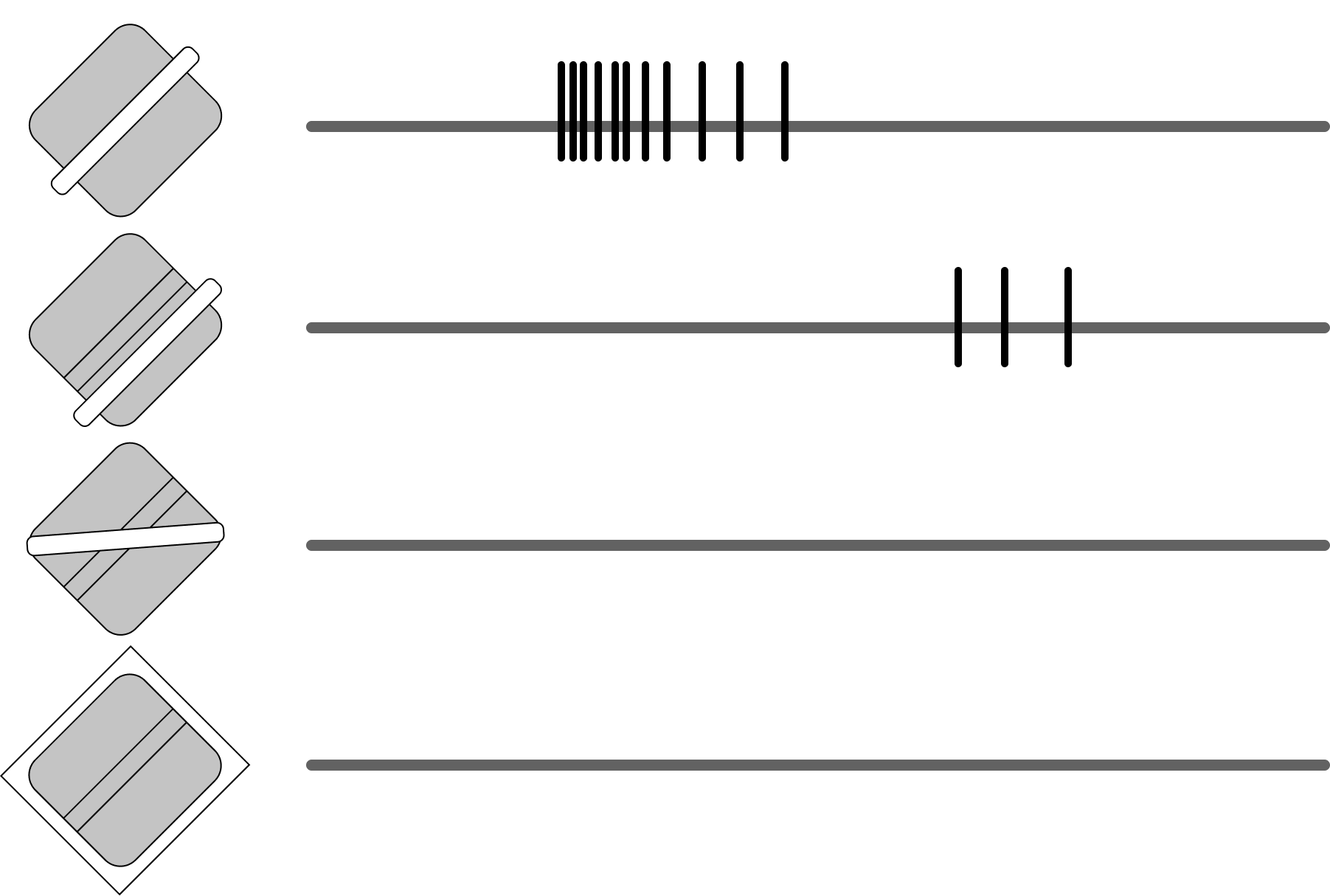}
\caption{The response of a simple cell with a visual receptive field, when stimulated with four different patterns. The first example shows a series of action potentials, also known as spike train, when stimulated with an input pattern with the optimum size, position, and orientation. The second example shows a response to an input pattern that has an optimal orientation and size, but a slightly unaligned position with the receptive field. The neuron emits fewer action potentials. In the third example, the orientation of the input pattern does not align at all with the receptive field of the neuron, causing the neuron not to respond. In the last example, the input pattern represents a large fully illuminated region, which does not align with the neuron's receptive field, and results in no firing.}
\label{fig:simple_cell1}
\end{figure}

Simple cells are neurons that fire whenever stimulated with a specific input pattern. In the case of visual neurons, simple cells specialize to activate to specifically oriented edges, or bars. These neurons are basic pattern detectors that activate only to one specific input (see figure \ref{fig:simple_cell1}). In other words, If the neuron "sees" its pattern, it will fire rapidly. If however the pattern does not fully align with its receptive field, the neuron fires less frequently, or not at all. The same principle of basic pattern detection functions for other types of receptive fields as well. Simple cells with auditory receptive fields for example respond selectively to sound frequencies.

\begin{figure}[h]
\centering\includegraphics[width=1.0\linewidth]{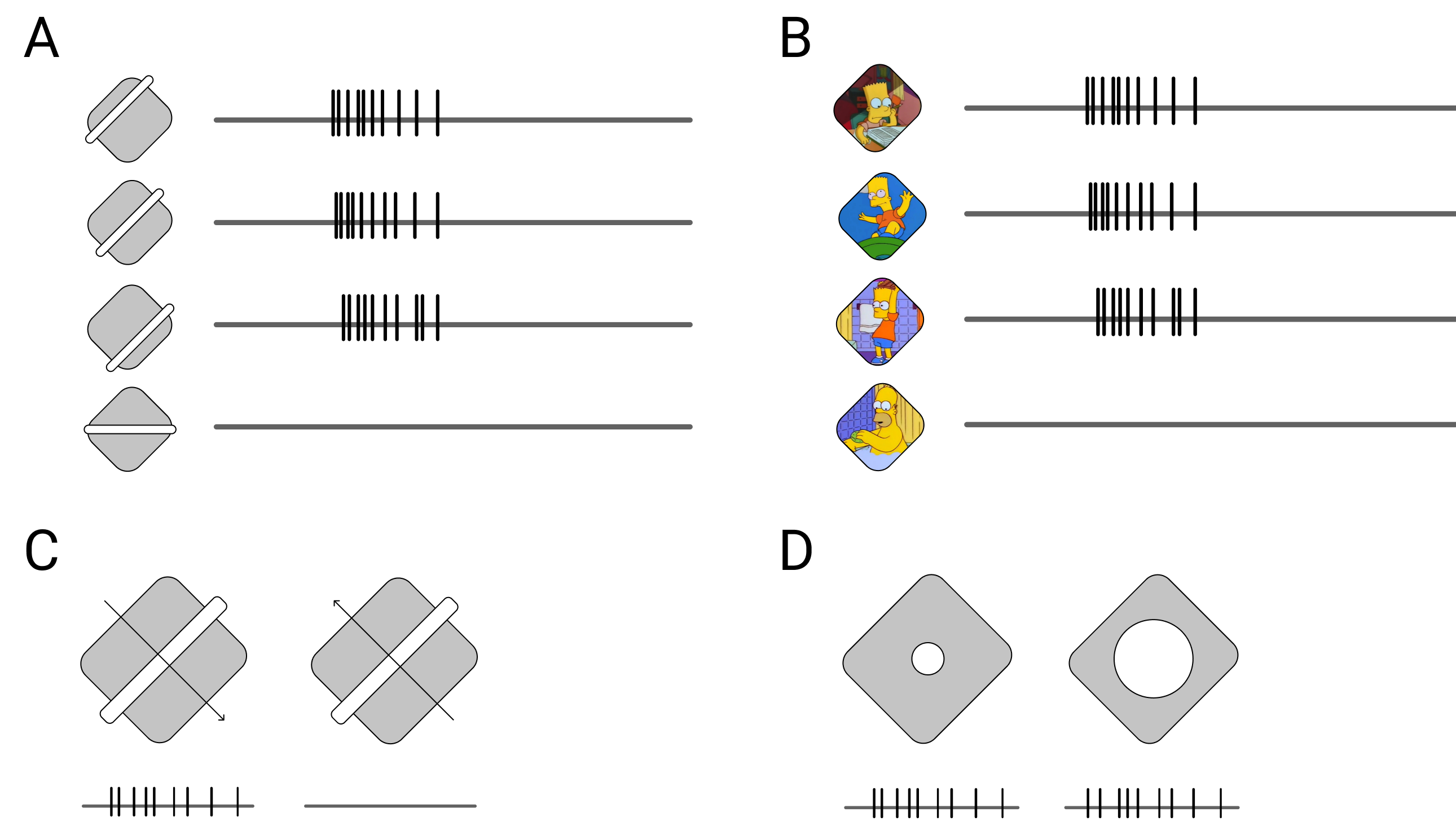}
\caption{\textbf{A} - The response of a complex cell with a visual receptive field, when stimulated with four different patterns. The receptive field of this complex cell is transition invariant. \textbf{B} - The response of a complex cell with a visual receptive field, when stimulated with four different Simpsons snapshots. The receptive field of this complex cell fires whenever Bart is inside the receptive field, but does not fire when Homer is presented. The receptive field of this complex cell has elements of viewpoint invariance. \textbf{C} - A visual receptive field with a directional preference. The neuron fires rapidly whenever its pattern moves across one direction, while remaining inactive if the same pattern moves in the other direction. \textbf{D} - A scale invariant visual receptive field. The neuron fires when a specific input pattern is used, regardless of the pattern's scale.}
\label{fig:complex_cell}
\end{figure}

Complex cells on the other hand, as the name suggests, have more complex receptive fields. In the case of visual receptive fields, some complex cells fire whenever stimulated with a specific input pattern, regardless of its position in the receptive field (see figure \ref{fig:complex_cell} \textit{A}). In other words, similarly to artificial convolutional neural networks (CNNs) \cite{fukushima_neocognitron_1980-1, lecun_backpropagation_1989}, the receptive field of these cells is said to have \textit{transition invariance}. Other complex cells might have a receptive field with a \textit{directional preference} (see figure \ref{fig:complex_cell} \textit{C}). These cells fire whenever stimulated with an input pattern that moves at a specific direction regardless of time, but do not fire if the same input pattern moves across the receptive field in the opposite direction. Similarly to artificial recurrent neural networks (RNNs) \cite{hopfield_neural_1982, hochreiter_long_1997}, the receptive field of these cells are said to have \textit{temporal invariance}. Some complex cells fire when stimulated with a specific pattern, regardless of the pattern's size inside the receptive field (see figure \ref{fig:complex_cell} \textit{D}). These complex cells have a receptive field that is \textit{scale invariant}. 

With every successive processing stage, the receptive field of a neuron grows in size and complexity. In the early processing stages, the receptive field of neurons is specialized towards very specific input patterns, as seen in simple cells. The receptive field of neurons in "deeper" processing stages start to form various invariant abilities, such as transition invariance, scale invariance, rotational invariance, etc., as seen in complex cells with visual receptive fields. At a given stage, the receptive field of complex cells can reach even more advanced abilities, such as recognizing objects regardless of viewing angle (see figure \ref{fig:complex_cell} \textit{B}). The receptive field of these complex cells are said to have \textit{viewpoint invariance}. How neurons form their given receptive field and the various kinds of invariances remains unclear. \textit{It seems as if the architecture of biological neural networks already contains the necessary inductive biases in its design, to enable the learning of generalized invariances}.

\subsection{Lateral inhibition}
\label{ss: lateral_inhibition}
Besides excitatory pyramidal neurons, the brain also contains a wide spectrum of specialized inhibitory interneurons. While pyramidal neurons are quite large, have an apical dendrite, and can form connections with distant regions of the brain, interneurons are smaller in size, do not have an apical dendrite, and usually form many local connections.

\begin{figure}[h]
\centering\includegraphics[width=0.9\linewidth]{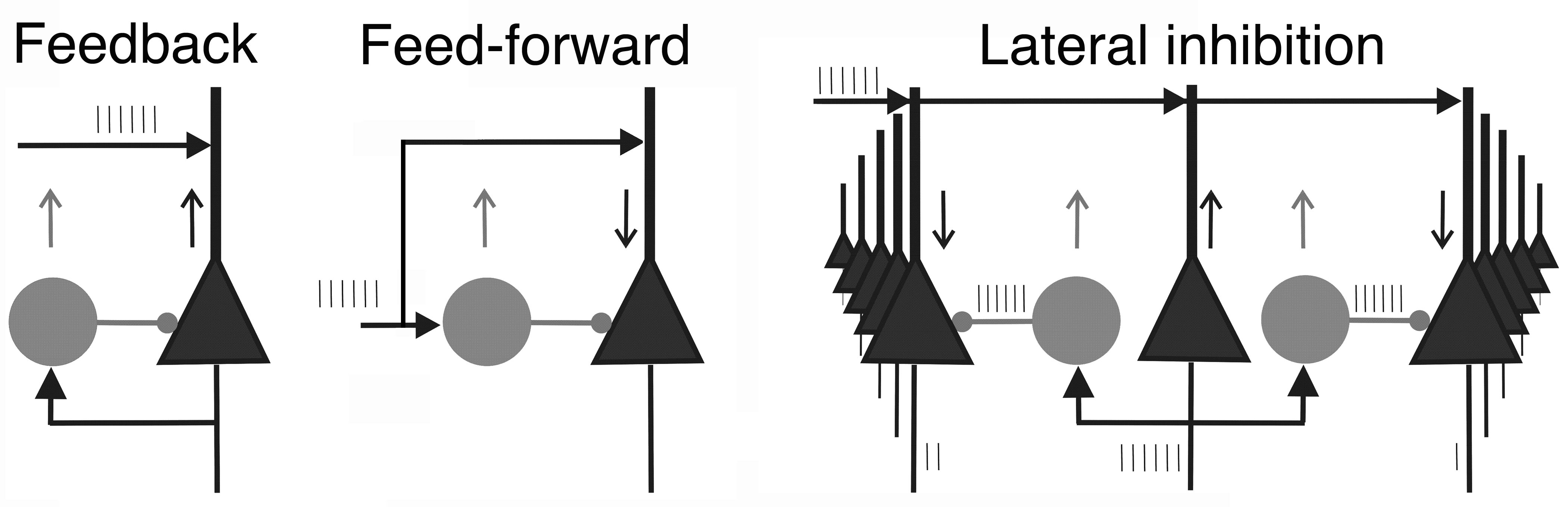}
\caption{The various network configurations of inhibition. Circles represent interneurons, whereas triangles represent pyramidal neurons. Upward arrows indicate the activation of a neuron, whereas downward arrows indicate the inhibition of a neuron. A sequence of vertical lines represents a spike train. In a \textbf{feedback network}, the activation of a pyramidal neuron $A$, results in the activation of an interneuron, which results in the inhibition of $A$. This network organization might serve as a regulatory mechanism for a neuron's firing pattern. In a \textbf{feed-forward network}, the activation of an interneuron results in the inhibition of another pyramidal neuron. A \textbf{lateral inhibitory configuration} occurs when several pyramidal neurons share one ore more common interneurons. In this network configuration, pyramidal neurons compete with one another by trying to fire before the other pyramidal neurons, thereby activating their common interneurons and inhibiting the rest. Illustration reprinted from \cite{prof_gyorgy_buzsaki_neural_nodate}}
\label{fig:forms_of_inhibition}
\end{figure}

Neural circuits containing both pyramidal neurons and interneurons can organize in several ways, giving rise to networks with complex properties (see figure \ref{fig:forms_of_inhibition}). In this work, we are interested in the organization that enables \textit{lateral inhibition}. Lateral inhibition occurs when a pyramidal neuron activates an inhibitory interneuron, which in turn suppresses the activity of surrounding pyramidal neurons. Networks with lateral inhibitory configurations (from now on simply referred to as "\textbf{competitive circuits}") form a "Winner take all" competition between the pyramidal neurons within an area. The neuron that fires before the other neurons "wins" the competition, by inhibiting the rest.

\subsection{Temporal codes}
\label{ss: temporal codes}

"\textit{Individual nerve cells were formerly thought to be unreliable, idiosyncratic, and incapable of performing complex tasks without acting in concert and thus  overcoming their individual errors. This was quite wrong, and we now realise their apparently erratic behaviour was caused by our ignorance, not the neuron's incompetence.}" \cite{barlow_single_1972}

\begin{figure}[h]
\centering\includegraphics[width=0.6\linewidth]{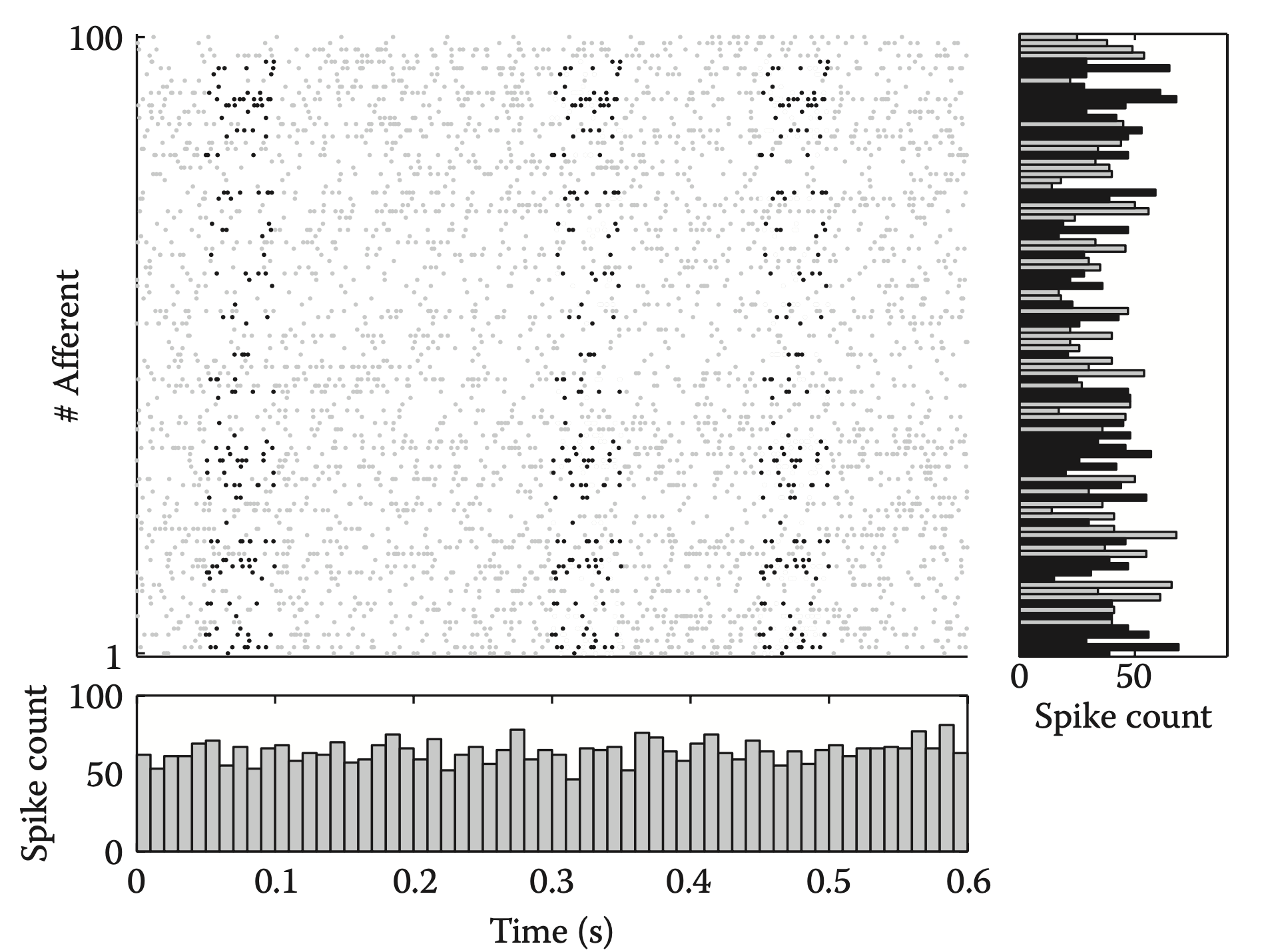}
\caption{The detection of spatiotemporal spike patterns. The Y-axis of the plot represents 100 neurons, 50 of which take part in a repeating precise spike pattern. The X-axis represents time. The dots within the plot are spikes. The darker dots represent spikes that are part of a precise spike pattern. The bottom panel plots the overall spike counts over 10 ms time intervals. The right panel plots the spike counts of each neuron over the whole period.
Reprinted from \cite{masquelier_learning_2013}. Modified from \cite{masquelier_spike_2008}}
\label{fig:spike_patterns}
\end{figure}

Neurons transmit and receive information via sequences of action potentials, also known as spike trains. We now know that the number of action potentials within spike trains is not the only variable used to encode information; the timing of the emitted spikes can also be part of the code \cite{mainen_reliability_1995, bohte_evidence_2004, gerstner_neural_1997, strong_entropy_1998, theunissen_temporal_1995}. Recent evidence supports the existence of precise spike patterns with millisecond-level precision (see figure \ref{fig:spike_patterns}) \cite{prut_spatiotemporal_1998, putney_precise_2019, thorpe_spike_1990, butts_temporal_2007, gollisch_rapid_2008}. These patterns have been found in the olfactory system \cite{egea-weiss_high_2018, bohte_evidence_2004}, auditory system \cite{decharms_primary_1996, bohte_evidence_2004}, visual system \cite{bialek_reading_1991, strong_entropy_1998}, somatosensory system \cite{panzeri_role_2001, mackevicius_millisecond_2012}, and within other sensory systems \cite{lawhern_spike_2011}.

The role and mechanism behind the precise spike patterns remains however enigmatic. Figure \ref{fig:spike_patterns} perfectly demonstrates the elusiveness of the problem: While looking at each neuron's spike train, the spikes within the spike trains seem completely random. At first glance, it looks as if there's no temporal information encoded within the spike trains. \textit{Only after performing an exhaustive search, do the repeating millisecond-level precise spike patterns become apparent} (represented as black dots).

\section{The Koha Model}
\label{s:the_hypothesis}
The previous section addressed some of the biological observations required for the Koha Model. This section on the other hand is speculative in nature. In this section, I present a model which aims to explain two phenomena:
\begin{enumerate}
    \item How information is processed and stored within neural circuits. 
    \item How neurons learn to become pattern detectors in an unsupervised way.
\end{enumerate}

\subsection{The Koha code}
\label{ss:the_koha_code}
We have seen in section \ref{ss: patterns in the substrate} that dendritic spines serve as electrical and biochemical compartments, able to amplify, as well as dampen incoming signals \cite{araya_sodium_2007, araya_activity-dependent_2014}. In fact we now know that most of the filtering of local potentials in a neuron happens within the spine neck and not along the dendrite \cite{araya_spine_2006}. The spine is somehow able to know when to let information pass through its neck and when not. The amplification, or dampening of incoming signals inside dendritic spines directly depends on the length and narrowness of the spine neck. Changes in the length and narrowness of the spine neck therefore, have a direct effect on the compartmentalization of the spine \cite{basu_role_2018}. We also know that due to the spine neck's highly dynamic actin cytoskeleton, dendritic spines can change their physical structure in seconds, following an appropriate stimulus \cite{yuste_mechanisms_1999, yuste_dendritic_1995, fischer_rapid_1998, dunaevsky_developmental_1999, korkotian_bidirectional_1999}. This change in shape is governed by actin-associated proteins such as drebrin \cite{hayashi_modulatory_1996}, $\alpha$-actin \cite{wyszynski_competitive_1997}, gelsolin \cite{furukawa_actin-severing_1997}, fodrin \cite{carlin_identification_1983}, etc. The properties of these proteins in turn depend on the calcium levels within a spine \cite{deller_potential_2000}. Increases in Intracellular calcium levels might therefore change the properties of microfilament-associated proteins, which then interact with the actin cytoskeleton of the spine, which results in the reshaping of the spine neck, which in turn influences the compartmentalization of the spine.

In short, if the dendritic spine could control its internal calcium levels in response to incoming signals, then it could also dampen, or amplify those signals by changing its shape. This is the main idea of the Koha model. Almost all mushroom spines possesses an enigmatic organelle known as a spine apparatus \cite{spacek_three-dimensional_1997}. Spine apparati are now known to be intracellular calcium stores that are able to quickly absorb, as well as release $Ca^{2+}$, depending on internal feedback \cite{fifkova_possible_1985, verkhratsky_physiology_2005, holbro_differential_2009}. A fascinating study showed that synaptic stimulation can cause the spine apparatus to release its internal calcium stores \cite{emptage_single_1999}. 
According to the proposed hypothesis, I argue that there is a molecular unit within the dendritic spine that encodes a temporal pattern in the form of a sequence of "on" states and "off" states. Whenever a dendritic spine receives a sequence of inputs, it will compare the spike train's temporal pattern, with its internal temporal code. If the received spike train's temporal pattern is similar to the spine's temporal code, information flows from the spine to its dendrite freely. If on the other hand the spike train's temporal pattern is significantly dissimilar from the spines internal code, the spine apparatus gets notified to increase intracellular calcium levels, which changes the shape of the spine and dampens the incoming signals. Signal processing on a local level is therefore controlled mechanically at the spine neck, the behavior of which depends on the temporal pattern of the incoming spike train.

\begin{figure}[h]
\centering\includegraphics[width=0.7\linewidth]{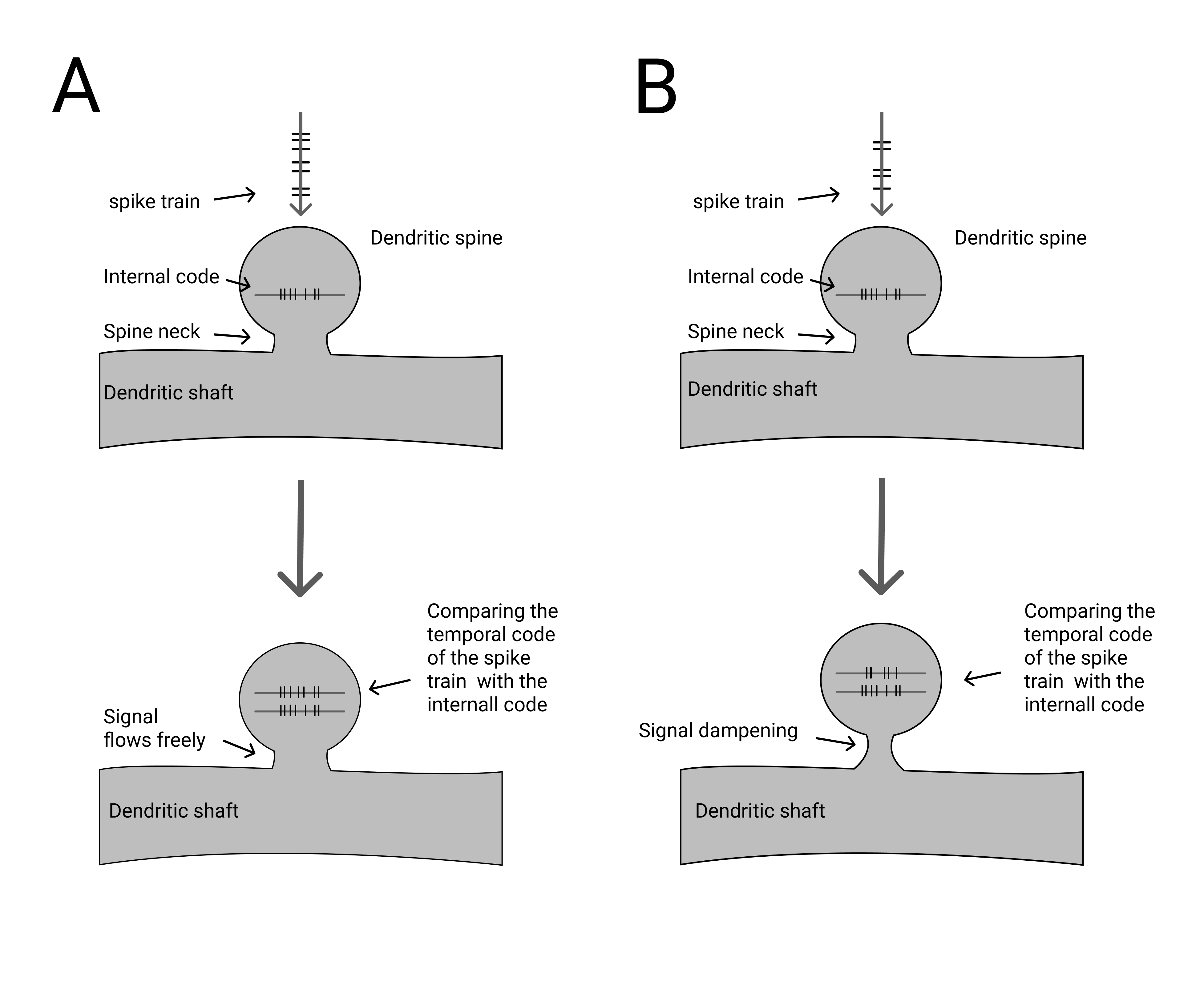}
\caption{\textbf{A} - The response of a dendritic spine when stimulated with a spike train whose temporal code is similar to that of the spine's internal code. The spine neck remains unchanged and the signal can pass freely. \textbf{B} - The response of a dendritic spine when stimulated with a spike train whose temporal code is not similar to that of the spine's internal code. The spine neck becomes longer and narrower, dampening the strength of the incoming signal in the process.}
\label{fig:koha_code}
\end{figure}

\subsection{Specialization through competition}
\label{ss:specialization_through_competition}
\begin{figure}[h]
\centering\includegraphics[width=0.8\linewidth]{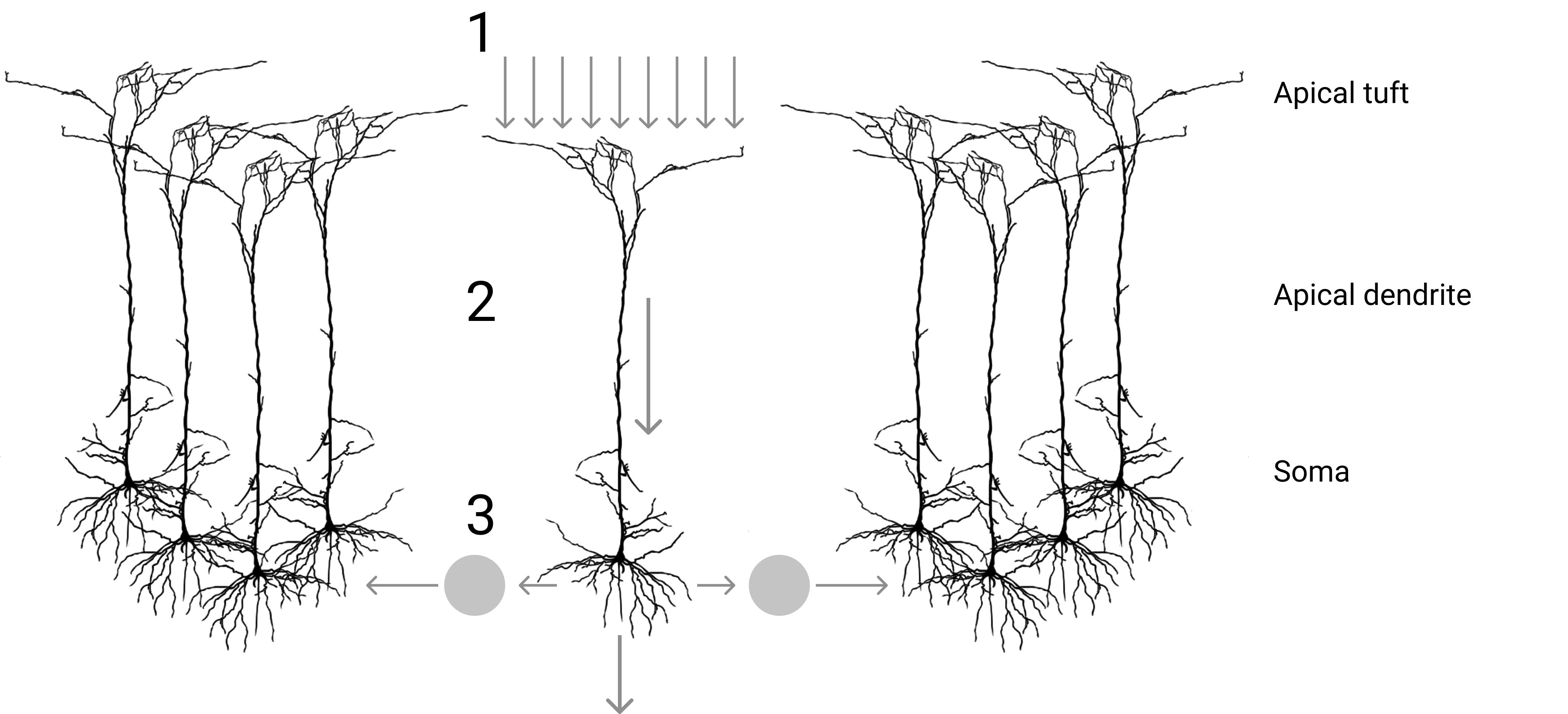}
\caption{The process of forward propagation within the Koha model. \textbf{1} - The dendritic spines of the apical tuft receives excitatory signals from a previous layer. Each spine computes the similarity between the temporal pattern of the input signal and that of its internal code (see section \ref{ss:the_koha_code}). \textbf{2} - The processed signals move from the apical tuft, through the apical dendrite to the soma. The signal of each neuron will have to cross this distance. \textbf{3} - The neuron whose propagating signals reach the soma first, will fire a series of action potentials, which activate surrounding inhibitory interneurons (shown as circles). The activated interneurons then inhibit all the other neurons within the competition.}
\label{fig:forward_propagation}
\end{figure}

To summarize, every dendritic spine on the apical dendride of a pyramidal neuron computes a distance measure between the temporal pattern of the input signal and that of its internal code. When a neuron receives inputs at its dendritic spines, it is essentially in a competition with all other neurons that receive the same inputs. Neurons with internal codes that are substantially different from the input signals attenuate the signals (by altering the spine necks of their spines) and delay the propagation of the signals from the apical tuft to the soma of the neuron (see figure \ref{fig:forward_propagation}). The neuron with the internal codes most similar to the input signals, will allow the signals to move freely from the apical dendrite to the soma. The "best matching neuron", that is, the neuron whose signals past through the apical dendrite to the soma the fastest, will fire before the other pyramidal neurons. If the neurons are organized as a competitive circuit (see section \ref{ss:lateral_inhibition}), the best matching neuron will activate neighboring inhibitory interneurons, which then inhibit the other pyramidal neurons in that participated in the competition. 

\begin{figure}[h]
\centering\includegraphics[width=0.8\linewidth]{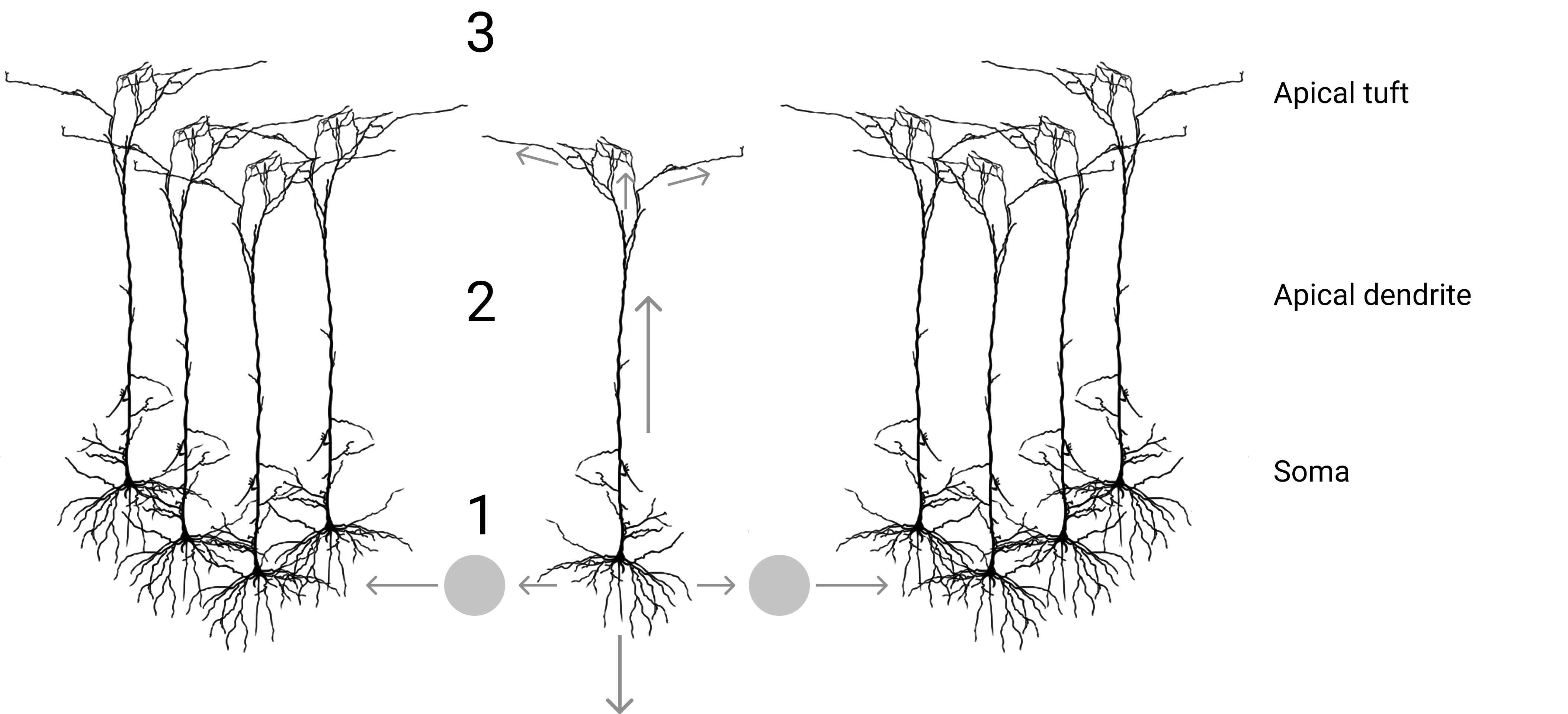}
\caption{The process of backward propagation within the Koha model. \textbf{1} - The winning neuron fires a series of action potentials to the next layer, as well as to its neighboring interneurons (shown as circles), which inhibit the surrounding neurons. In parallel to that, the winning neuron fires a series of backpropagating action potentials in the direction of its apical dendrite. \textbf{2} - The propagating signals move from the winning neuron's soma, through its apical dendrite, all the way to the individual dendritic spines at the apical tuft. \textbf{3} - Each dendritic spine that participated in the competition updates its internal code, so as to become slightly more similar to the temporal pattern of the spike train that each dendritic spine received (see section \ref{ss:the_koha_code}).}
\label{fig:backward_propagation}
\end{figure}

We now know that whenever a neuron fires an action potential, another impulse from the soma is also generated that propagates backwards through the apical dendrite of the neuron. This process is also known as Neural backpropagation. When a backpropagating action potential contacts previously activated dendritic spines, a superlinear rise in internal calcium levels occurs inside the dendritic spines. \cite{yuste_dendritic_1995, yuste_mechanisms_1999}. Neighboring spines that were not activated prior to the backpropagating signal would be unaffected. I argue that the backpropagating action potential can be viewed as a mechanism to inform dendritic spines that their action resulted in a best matching neuron. That is, it informs the dendritic spines down the apical dendrite that their neuron won the competition (see figure \ref{fig:backward_propagation}). Every dendritic spine will then modify their internal code to become slightly more similar to the temporal code that they received during the competition. In a way the backpropagating signal can be seen as a message to all the dendritic spines - "We won! Update your codes slightly so that we can win the next time even faster!" This process can be viewed as a form of \textit{competitive learning} \cite{rumelhart_parallel_1986}, which is a variant of Hebbian learning \cite{hebb_organization_1949}. The internal codes of each neurons' dendritic spines change over time, becoming in the process more specialized towards a specific input. This in turn makes it more likely for a neuron to win the next competition again, if the same given input is provided. 

It is worth noting that the model so far explains several organizational and morphological observations: We know that dendritic spines receive most of the excitatory inputs in neurons, and that practically every dendritic spine has an excitatory synapse on its head \cite{gray_electron_1959, arellano_non-synaptic_2007, colonnier_synaptic_1968}. Excitatory axons prefer to terminate on spines and almost never connect directly to dendritic shafts (the body of the dendrite) \cite{colonnier_synaptic_1968}. This consistent observation makes sense, if we assume that dendritic spines serve as computational units that try to find specific temporal patterns within excitatory inputs. Inhibitory axons on the other hand almost always connect directly to dendritic shafts, rather than on spines \cite{somogyi_salient_1998}. This also makes sense, if we assume that one of the roles of inhibitory neurons is to signal the end of a neural competitions. Connecting directly onto dendritic shafts allows for faster propagation, than propagating an inhibitory signal through the spine neck. It also explains why most observed interneurons are essentially spineless \cite{yuste_dendritic_2010}. In the Koha model, most inhibitory neurons serve as signal relayers, not as information processing units. The competitive circuit also explains why pyramidal neurons posses about 10 times more local connections with fast-spiking inhibitory neurons, than with other pyramidal neurons, with most of the connections being reciprocal (in the layer 2/3 of the rat neocortex at least) \cite{holmgren_pyramidal_2003}.

\section{The Koha model as an associative memory network}
The Koha model can be viewed as an associative memory model. The idea of an associative memory network is to map an input to the most similar internal pattern. Machine learning models such as the Hopfield Network \cite{hopfield_neural_1982}, and the self organizing map \cite{kohonen_self-organized_1982} are classical associative memory models. Recently self-attention-based architectures, such as Transformers \cite{vaswani_attention_2017} have gained wide popularity in the machine learning space. The newer Modern Hopfield Network \cite{ramsauer_hopfield_2021}, an artificial network that uses the Transofmer's attention mechanism as an update rule, is of particular interest. The Modern Hopfield Network shares many parallels with the Koha hypothesis of biological memory and could be considered as a continuous implementation of the biological hypothesis. The "HopfieldLayer" implementation of the Modern Hopfield Network can be defined as:

\begin{equation}
\label{eq:hopfield}
\bm{Z} = softmax \left( \beta \bm{R} \bm{W^T_K} \right) \bm{W_V}
\end{equation}

The network stores learnable patterns in a weight matrix $\bm{W_K}$, also referred to as the "key weight matrix". The matrix $\bm{R}$ is the input set, where every row represents a data point. $\beta$ is a parameter for controlling the "temperature" of the model, whereas $\bm{W_V}$ is the "value weight matrix".

\begin{figure}[h]
\centering\includegraphics[width=0.8\linewidth]{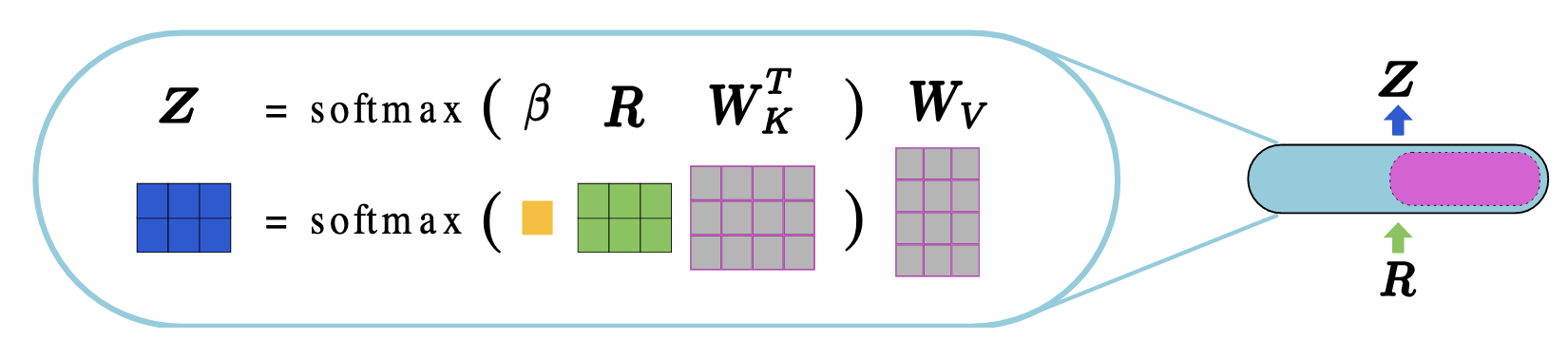}
\caption{The "HopfieldLayer" configuration of the Modern Hopfield Network. Reprinted from \cite{ramsauer_hopfield_2021}}
\label{fig:hopfield}
\end{figure}

From a biological perspective, we can imagine every column of $\bm{W_K}$ to be a neuron with its temporal codes representing the values of the column. When computing the dot product of a normalized input $\bm{R_i}$ with the key weight matrix $\bm{W_K}$, we essentially compute the cosine similarity between the input and every pattern of the weight matrix. The resulting vector of cosine similarities represents a score that tells us which neuron is more similar to the given input. Passing the computed scores to the softmax function, creates a probability distribution in which patterns more similar to the input are closer to 1, and dissimilar patterns closer to 0. The described process can be viewed as a neural competition. $W_V$ represents the output spike train of the neurons, which in themselves would contain precise spike patterns. If dendritic spines truly posses the ability to scan for precise spike patterns, than it is natural to assume that every neuron has a specific temporal "ID", that is, a spike pattern that it transmits. If the activity of presynaptic neuron is too much out of sync with its original "temporal ID", the dendritic spines of postsynaptic neurons will dampen the signal, and therefore weaken the presynaptic neuron's vote in the dendritic computation of the postsynaptic neuron's temporal pattern. In fact, an ingenious experiment by \cite{oesch_direction-selective_2005} showed that the temporal order of a spike train emitted by a neuron, does not get computed in the soma, or even at the axonal initial segment, but rather in the dendritic tree of the neuron.

\section{Discussion, Criticism, \& Future Work}
\label{s:discussions}
The current model proposes a mechanism for dendridic spines to scan incoming signals for temporal patterns, But the exact molecular mechanism remains unexplained. Synaptopodin's preferential location at the spine neck, and its close association with the spine apparatus and the actin cytoskeleton, suggests that synaptopodin is somehow involved in the rapid plasticity of the spine's actin cytoskeleton. One study suggests that synaptopodin could provide a link between the actin cytoskeleton of the spine neck, and the membrane of the spine apparatus \cite{deller_potential_2000}. Considering that some membrane-bound receptors have mechanosensitive properties \cite{paoletti_mechanosensitivity_1994}, a mechanical interaction between synaptopodin and the receptor molecules of the spine apparatus might be possible. If correct, calcium release of the spine apparatus might be determined by the spine apparatus' interactions with synaptopodin. 

Another detail not explain in depth, is the role of thin spines. Thin spines in contrast to most mushroom spines do not have a spine apparatus. Because of this fact, one might argue that the spine apparatus cannot be the only mechanism for how dendritic spines filter information. Which is correct, calcium gates within the spine membrane play certainly an important role. It is however important to note that thin spines have significantly longer necks than mushroom spines, and are as a result biochemically more isolated. Thin spines might therefore not play as much of a role in dendritic computations as mushroom spines do. I argue that in the "neural democracy" of the dendritic tree, the "vote" of mushroom spines has more value than that of a thin spine. Considering that mushroom spines can remain stable for a lifetime, while thin spines can appear and disappear throughout life \cite{bourne_thin_2007, holtmaat_transient_2005}, it would make sense to weigh the contribution of mushroom spines more.

\begin{figure}[h]
\centering\includegraphics[width=0.7\linewidth]{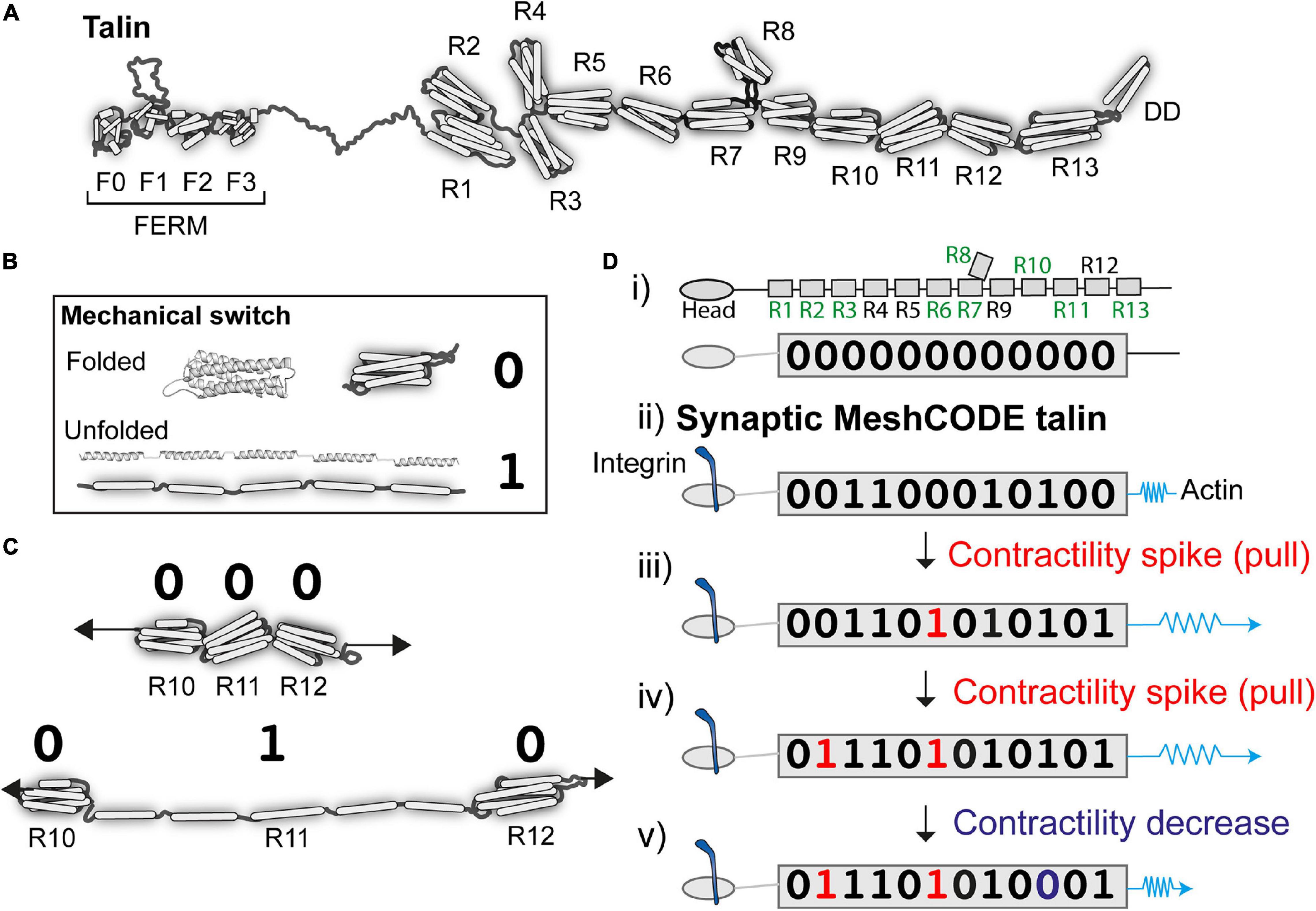}
\caption{}
\label{fig:talin}
\end{figure}

The Koha model also argues for the existence of a code within dendritic spines, but it does not describe the physical structure of the hypothetical code. There are however good candidates. The talin protein, a highly conserved \cite{senetar_gene_2005} mechanosensitive cytoskeletal protein in vertebrates, could be encoding the temporal information that the Koha model suggests exists. In an excellent paper \cite{goult_mechanical_2021}, Goult proposes the MeshCODE hypothesis, in which he argues that talin has the right properties to serve as a memory molecule. The talin protein consist of 13 "switches" which can reside in two thermodynamically stable states. The MeshCODE hypothesis refers to these switch states as "on" an "off" states and proposes that information can be encoded on the molecule by either folding, or unfolding the talin switches (see figure \ref{fig:talin}). Considering that both talin and synaptopodin are tightly connected to the actin of the cytoskeleton, a switch controlled mechanical interaction between talin, actin, and synaptopodin might exist, and could trigger the release of the spine apparatus' calcium store down the line. Note however that this is a hypothesis build on top of another hypothesis and that the correct answer might be something entirely else. 

Another missing component of the Koha model, is the formation of complex cells. The described competitive learning process can give rise to simple cells with simple receptive fields, and even to cells with directional preferences (by adding skip-connections), but it does not describe, or propose a biologically plausible way for the formation of "generalized invariances". That is, it does not describe how neurons become complex cells with invariant abilities, such as viewpoint invariance. There are several existing hypotheses for how complex cells form \cite{martinez_complex_2003}, but none seem compelling enough on a biological level, that is, none show a concrete process for how invariances are learned naturally. One component that has been purposefully put on the side in this work is the role of basal dendrites. The whole design of the proposed competitive learning process depends on the apical dendrite and its dendritic spines, but ignored the role of basal dendrites, which are in fact just as important. I argue that by incorporating a few more elements to the Koha model (such as basal dendrites and the nonliniarities introduced in the dendritic processing of apical dendrites), one can explain and model the formation of generalized invariances as well. This is also why the optimization criteria in the Modern Hopfield Network example was not covered in detail. The topic of generalized invariances deserves and requires a paper in itself. How biological neural networks learn invariances, and how generalized invariances can be used to design better performing artificial neural networks will be the focus of my subsequent paper.

\section{Conclusion}
This work introduces two novel ideas with widespread implications for the field of neuroscience:

On a micro level, it argues for the existence of a temporal code within every dendritic spine. Just as genes are the molecular unit of heredity, the Koha code is argued to be the unit of memory.  It shows how dendritic spines use these temporal codes to scan for precise spike patterns in their synaptic inputs. It describes a biologically plausible process for how signal filtration occurs within spine necks, as well as provides compelling evidence for the existence of such a mechanism within dendritic spines.

On a macro level, it explains how neurons within competitive circuits can learn to become pattern detectors, through competitive learning. In this model, the chance of a neuron to become the "winning neuron" within a competitive circuit, directly depends on the temporal codes within a neuron's dendritic spines.

\section{Acknowledgements}
I would like to thank Taulant Ramabaja and Rita Qarkaxhija for their support, feedback, and stimulating discussions; without you the ideas in this work would have never been written. A debt of gratitude is also owed to everyone at the Institute for Machine Learning at the Johannes Kepler University (JKU) in Linz. The machine learning theory taught by the professors and researchers at the JKU served as an invaluable compass in the research and writing of this paper.

\bibliographystyle{unsrt}  
\bibliography{references}  %%% Remove comment to use the external .bib file (using bibtex).

\begin{thebibliography}{10}

\bibitem{holmgren_pyramidal_2003}
Carl Holmgren, Tibor Harkany, Björn Svennenfors, and Yuri Zilberter.
\newblock Pyramidal cell communication within local networks in layer 2/3 of
  rat neocortex.
\newblock {\em The Journal of Physiology}, 551(Pt 1):139--153, August 2003.

\bibitem{henze_dendritic_1996}
D.~A. Henze, W.~E. Cameron, and G.~Barrionuevo.
\newblock Dendritic morphology and its effects on the amplitude and rise-time
  of synaptic signals in hippocampal {CA3} pyramidal cells.
\newblock {\em The Journal of Comparative Neurology}, 369(3):331--344, June
  1996.

\bibitem{spruston_pyramidal_2008}
Nelson Spruston.
\newblock Pyramidal neurons: dendritic structure and synaptic integration.
\newblock {\em Nature Reviews. Neuroscience}, 9(3):206--221, March 2008.

\bibitem{yuste_dendritic_2010}
Rafael Yuste.
\newblock {\em Dendritic {Spines}}.
\newblock 2010.
\newblock Publisher: The MIT Press.

\bibitem{peters_small_1970}
Alan Peters and Ita~R. Kaiserman‐Abramof.
\newblock The small pyramidal neuron of the rat cerebral cortex. {The}
  perikaryon, dendrites and spines.
\newblock {\em American Journal of Anatomy}, 127(4):321--355, 1970.
\newblock \_eprint:
  https://onlinelibrary.wiley.com/doi/pdf/10.1002/aja.1001270402.

\bibitem{harris_three-dimensional_1992}
K.~M. Harris, F.~E. Jensen, and B.~Tsao.
\newblock Three-dimensional structure of dendritic spines and synapses in rat
  hippocampus ({CA1}) at postnatal day 15 and adult ages: implications for the
  maturation of synaptic physiology and long-term potentiation.
\newblock {\em The Journal of Neuroscience: The Official Journal of the Society
  for Neuroscience}, 12(7):2685--2705, July 1992.

\bibitem{berry_spine_2017}
Kalen~P. Berry and Elly Nedivi.
\newblock Spine {Dynamics}: {Are} {They} {All} the {Same}?
\newblock {\em Neuron}, 96(1):43--55, September 2017.

\bibitem{bourne_thin_2007}
Jennifer Bourne and Kristen~M. Harris.
\newblock Do thin spines learn to be mushroom spines that remember?
\newblock {\em Current Opinion in Neurobiology}, 17(3):381--386, June 2007.

\bibitem{holtmaat_transient_2005}
Anthony J. G.~D. Holtmaat, Joshua~T. Trachtenberg, Linda Wilbrecht, Gordon~M.
  Shepherd, Xiaoqun Zhang, Graham~W. Knott, and Karel Svoboda.
\newblock Transient and persistent dendritic spines in the neocortex in vivo.
\newblock {\em Neuron}, 45(2):279--291, January 2005.

\bibitem{gray_electron_1959}
E.~G. Gray.
\newblock Electron {Microscopy} of {Synaptic} {Contacts} on {Dendrite} {Spines}
  of the {Cerebral} {Cortex}.
\newblock {\em Nature}, 183(4675):1592--1593, June 1959.
\newblock Number: 4675 Publisher: Nature Publishing Group.

\bibitem{arellano_non-synaptic_2007}
J.~I. Arellano, A.~Espinosa, A.~Fairén, R.~Yuste, and J.~DeFelipe.
\newblock Non-synaptic dendritic spines in neocortex.
\newblock {\em Neuroscience}, 145(2):464--469, March 2007.

\bibitem{colonnier_synaptic_1968}
M.~Colonnier.
\newblock Synaptic patterns on different cell types in the different laminae of
  the cat visual cortex. {An} electron microscope study.
\newblock {\em Brain Research}, 9(2):268--287, July 1968.

\bibitem{somogyi_salient_1998}
Peter Somogyi, Gábor Tamás, Rafael Lujan, and Eberhard~H. Buhl.
\newblock Salient features of synaptic organisation in the cerebral
  {cortex1Published} on the {World} {Wide} {Web} on 3 {March} 1998.1.
\newblock {\em Brain Research Reviews}, 26(2):113--135, May 1998.

\bibitem{volfovsky_geometry_1999}
N.~Volfovsky, H.~Parnas, M.~Segal, and E.~Korkotian.
\newblock Geometry of dendritic spines affects calcium dynamics in hippocampal
  neurons: theory and experiments.
\newblock {\em Journal of Neurophysiology}, 82(1):450--462, July 1999.

\bibitem{wickens_electrically_1988}
J.~Wickens.
\newblock Electrically coupled but chemically isolated synapses: dendritic
  spines and calcium in a rule for synaptic modification.
\newblock {\em Progress in Neurobiology}, 31(6):507--528, 1988.

\bibitem{lisman_mechanism_1989}
J.~Lisman.
\newblock A mechanism for the {Hebb} and the anti-{Hebb} processes underlying
  learning and memory.
\newblock {\em Proceedings of the National Academy of Sciences of the United
  States of America}, 86(23):9574--9578, December 1989.

\bibitem{koch_function_1993}
C.~Koch and A.~Zador.
\newblock The function of dendritic spines: devices subserving biochemical
  rather than electrical compartmentalization.
\newblock {\em The Journal of Neuroscience: The Official Journal of the Society
  for Neuroscience}, 13(2):413--422, February 1993.

\bibitem{holmes_is_1990}
William~R. Holmes.
\newblock Is the function of dendritic spines to concentrate calcium?
\newblock {\em Brain Research}, 519(1):338--342, June 1990.

\bibitem{basu_role_2018}
Sreetama Basu and Raphael Lamprecht.
\newblock The {Role} of {Actin} {Cytoskeleton} in {Dendritic} {Spines} in the
  {Maintenance} of {Long}-{Term} {Memory}.
\newblock {\em Frontiers in Molecular Neuroscience}, 11, May 2018.

\bibitem{araya_spine_2006}
Roberto Araya, Jiang Jiang, Kenneth~B. Eisenthal, and Rafael Yuste.
\newblock The spine neck filters membrane potentials.
\newblock {\em Proceedings of the National Academy of Sciences of the United
  States of America}, 103(47):17961--17966, November 2006.

\bibitem{araya_sodium_2007}
Roberto Araya, Volodymyr Nikolenko, Kenneth~B. Eisenthal, and Rafael Yuste.
\newblock Sodium channels amplify spine potentials.
\newblock {\em Proceedings of the National Academy of Sciences of the United
  States of America}, 104(30):12347--12352, July 2007.

\bibitem{yuste_mechanisms_1999}
Rafael Yuste, Ania Majewska, Sydney~S. Cash, and Winfried Denk.
\newblock Mechanisms of {Calcium} {Influx} into {Hippocampal} {Spines}:
  {Heterogeneity} among {Spines}, {Coincidence} {Detection} by {NMDA}
  {Receptors}, and {Optical} {Quantal} {Analysis}.
\newblock {\em Journal of Neuroscience}, 19(6):1976--1987, March 1999.
\newblock Publisher: Society for Neuroscience Section: ARTICLE.

\bibitem{yuste_dendritic_1995}
Rafael Yuste and Winfried Denk.
\newblock Dendritic spines as basic functional units of neuronal integration.
\newblock {\em Nature}, 375(6533):682--684, June 1995.
\newblock Number: 6533 Publisher: Nature Publishing Group.

\bibitem{fischer_rapid_1998}
M.~Fischer, S.~Kaech, D.~Knutti, and A.~Matus.
\newblock Rapid actin-based plasticity in dendritic spines.
\newblock {\em Neuron}, 20(5):847--854, May 1998.

\bibitem{dunaevsky_developmental_1999}
Anna Dunaevsky, Ayumu Tashiro, Ania Majewska, Carol Mason, and Rafael Yuste.
\newblock Developmental regulation of spine motility in the mammalian central
  nervous system.
\newblock {\em Proceedings of the National Academy of Sciences of the United
  States of America}, 96(23):13438--13443, November 1999.

\bibitem{korkotian_bidirectional_1999}
E.~Korkotian and M.~Segal.
\newblock Bidirectional regulation of dendritic spine dimensions by glutamate
  receptors.
\newblock {\em Neuroreport}, 10(13):2875--2877, September 1999.

\bibitem{noguchi_spine-neck_2005}
Jun Noguchi, Masanori Matsuzaki, Graham~C.R. Ellis-Davies, and Haruo Kasai.
\newblock Spine-{Neck} {Geometry} {Determines} {NMDA} {Receptor}-{Dependent}
  {Ca2}+ {Signaling} in {Dendrites}.
\newblock {\em Neuron}, 46(4):609--622, May 2005.

\bibitem{araya_activity-dependent_2014}
Roberto Araya, Tim~P. Vogels, and Rafael Yuste.
\newblock Activity-dependent dendritic spine neck changes are correlated with
  synaptic strength.
\newblock {\em Proceedings of the National Academy of Sciences of the United
  States of America}, 111(28):E2895--2904, July 2014.

\bibitem{bar_periodic_2016}
Julia Bär, Oliver Kobler, Bas van Bommel, and Marina Mikhaylova.
\newblock Periodic {F}-actin structures shape the neck of dendritic spines.
\newblock {\em Scientific Reports}, 6(1):37136, November 2016.
\newblock Number: 1 Publisher: Nature Publishing Group.

\bibitem{spacek_three-dimensional_1997}
Josef Spacek and Kristen~M. Harris.
\newblock Three-{Dimensional} {Organization} of {Smooth} {Endoplasmic}
  {Reticulum} in {Hippocampal} {CA1} {Dendrites} and {Dendritic} {Spines} of
  the {Immature} and {Mature} {Rat}.
\newblock {\em Journal of Neuroscience}, 17(1):190--203, January 1997.
\newblock Publisher: Society for Neuroscience Section: Articles.

\bibitem{fifkova_possible_1985}
E.~Fifková.
\newblock A possible mechanism of morphometric changes in dendritic spines
  induced by stimulation.
\newblock {\em Cellular and Molecular Neurobiology}, 5(1-2):47--63, June 1985.

\bibitem{korkotian_fast_1998}
E.~Korkotian and M.~Segal.
\newblock Fast confocal imaging of calcium released from stores in dendritic
  spines.
\newblock {\em The European Journal of Neuroscience}, 10(6):2076--2084, June
  1998.

\bibitem{sharp_differential_1993}
A.~H. Sharp, P.~S. McPherson, T.~M. Dawson, C.~Aoki, K.~P. Campbell, and S.~H.
  Snyder.
\newblock Differential immunohistochemical localization of inositol
  1,4,5-trisphosphate- and ryanodine-sensitive {Ca2}+ release channels in rat
  brain.
\newblock {\em The Journal of Neuroscience: The Official Journal of the Society
  for Neuroscience}, 13(7):3051--3063, July 1993.

\bibitem{verkhratsky_physiology_2005}
Alexei Verkhratsky.
\newblock Physiology and pathophysiology of the calcium store in the
  endoplasmic reticulum of neurons.
\newblock {\em Physiological Reviews}, 85(1):201--279, January 2005.

\bibitem{holbro_differential_2009}
Niklaus Holbro, Åsa Grunditz, and Thomas~G. Oertner.
\newblock Differential distribution of endoplasmic reticulum controls
  metabotropic signaling and plasticity at hippocampal synapses.
\newblock {\em Proceedings of the National Academy of Sciences of the United
  States of America}, 106(35):15055--15060, September 2009.

\bibitem{deller_actin-associated_2000}
T.~Deller, T.~Merten, S.~U. Roth, P.~Mundel, and M.~Frotscher.
\newblock Actin-associated protein synaptopodin in the rat hippocampal
  formation: localization in the spine neck and close association with the
  spine apparatus of principal neurons.
\newblock {\em The Journal of Comparative Neurology}, 418(2):164--181, March
  2000.

\bibitem{deller_synaptopodin-deficient_2003}
Thomas Deller, Martin Korte, Sophie Chabanis, Alexander Drakew, Herbert
  Schwegler, Giulia~Good Stefani, Aimee Zuniga, Karin Schwarz, Tobias
  Bonhoeffer, Rolf Zeller, Michael Frotscher, and Peter Mundel.
\newblock Synaptopodin-deficient mice lack a spine apparatus and show deficits
  in synaptic plasticity.
\newblock {\em Proceedings of the National Academy of Sciences of the United
  States of America}, 100(18):10494--10499, September 2003.

\bibitem{deller_potential_2000}
T.~Deller, P.~Mundel, and M.~Frotscher.
\newblock Potential role of synaptopodin in spine motility by coupling actin to
  the spine apparatus.
\newblock {\em Hippocampus}, 10(5):569--581, 2000.

\bibitem{benedeczky_cisternal_1994}
I.~Benedeczky, E.~Molnár, and P.~Somogyi.
\newblock The cisternal organelle as a {Ca}(2+)-storing compartment associated
  with {GABAergic} synapses in the axon initial segment of hippocampal
  pyramidal neurones.
\newblock {\em Experimental Brain Research}, 101(2):216--230, 1994.

\bibitem{bas_orth_loss_2007}
Carlos Bas~Orth, Christian Schultz, Christian~M. Müller, Michael Frotscher,
  and Thomas Deller.
\newblock Loss of the cisternal organelle in the axon initial segment of
  cortical neurons in synaptopodin-deficient mice.
\newblock {\em The Journal of Comparative Neurology}, 504(5):441--449, October
  2007.

\bibitem{hubel_receptive_1959}
D.~H. Hubel and T.~N. Wiesel.
\newblock Receptive fields of single neurones in the cat's striate cortex.
\newblock {\em The Journal of Physiology}, 148(3):574--591, October 1959.

\bibitem{hubel_receptive_1968}
D.~H. Hubel and T.~N. Wiesel.
\newblock Receptive fields and functional architecture of monkey striate
  cortex.
\newblock {\em The Journal of Physiology}, 195(1):215--243, March 1968.

\bibitem{fukushima_neocognitron_1980-1}
Kunihiko Fukushima.
\newblock Neocognitron: {A} self-organizing neural network model for a
  mechanism of pattern recognition unaffected by shift in position.
\newblock {\em Biological Cybernetics}, 36(4):193--202, April 1980.

\bibitem{lecun_backpropagation_1989}
Y.~LeCun, B.~Boser, J.~S. Denker, D.~Henderson, R.~E. Howard, W.~Hubbard, and
  L.~D. Jackel.
\newblock Backpropagation {Applied} to {Handwritten} {Zip} {Code}
  {Recognition}.
\newblock {\em Neural Computation}, 1(4):541--551, December 1989.
\newblock Conference Name: Neural Computation.

\bibitem{hopfield_neural_1982}
J.~J. Hopfield.
\newblock Neural networks and physical systems with emergent collective
  computational abilities.
\newblock {\em Proceedings of the National Academy of Sciences},
  79(8):2554--2558, April 1982.
\newblock Publisher: National Academy of Sciences Section: Research Article.

\bibitem{hochreiter_long_1997}
Sepp Hochreiter and Jürgen Schmidhuber.
\newblock Long {Short}-{Term} {Memory}.
\newblock {\em Neural Computation}, 9(8):1735--1780, November 1997.

\bibitem{prof_gyorgy_buzsaki_neural_nodate}
Prof.~Gyorgy Buzsaki and {Dr. Peter Jonas}.
\newblock Neural inhibition - {Scholarpedia}.

\bibitem{barlow_single_1972}
H.~B. Barlow.
\newblock Single units and sensation: a neuron doctrine for perceptual
  psychology?
\newblock {\em Perception}, 1(4):371--394, 1972.

\bibitem{mainen_reliability_1995}
Z.~F. Mainen and T.~J. Sejnowski.
\newblock Reliability of spike timing in neocortical neurons.
\newblock {\em Science}, 268(5216):1503--1506, June 1995.
\newblock Publisher: American Association for the Advancement of Science
  Section: Reports.

\bibitem{bohte_evidence_2004}
Sander~M. Bohte.
\newblock The evidence for neural information processing with precise
  spike-times: {A} survey.
\newblock {\em Natural Computing}, 3(2):195--206, June 2004.

\bibitem{gerstner_neural_1997}
W.~Gerstner, A.~K. Kreiter, H.~Markram, and A.~V. Herz.
\newblock Neural codes: firing rates and beyond.
\newblock {\em Proceedings of the National Academy of Sciences of the United
  States of America}, 94(24):12740--12741, November 1997.

\bibitem{strong_entropy_1998}
S.~P. Strong, Roland Koberle, Rob~R. de~Ruyter~van Steveninck, and William
  Bialek.
\newblock Entropy and {Information} in {Neural} {Spike} {Trains}.
\newblock {\em Physical Review Letters}, 80(1):197--200, January 1998.
\newblock Publisher: American Physical Society.

\bibitem{theunissen_temporal_1995}
F.~Theunissen and J.~P. Miller.
\newblock Temporal encoding in nervous systems: a rigorous definition.
\newblock {\em Journal of Computational Neuroscience}, 2(2):149--162, June
  1995.

\bibitem{prut_spatiotemporal_1998}
Y.~Prut, E.~Vaadia, H.~Bergman, I.~Haalman, H.~Slovin, and M.~Abeles.
\newblock Spatiotemporal structure of cortical activity: properties and
  behavioral relevance.
\newblock {\em Journal of Neurophysiology}, 79(6):2857--2874, June 1998.

\bibitem{putney_precise_2019}
Joy Putney, Rachel Conn, and Simon Sponberg.
\newblock Precise timing is ubiquitous, consistent, and coordinated across a
  comprehensive, spike-resolved flight motor program.
\newblock {\em Proceedings of the National Academy of Sciences},
  116(52):26951--26960, December 2019.

\bibitem{thorpe_spike_1990}
Simon Thorpe.
\newblock Spike arrival times: {A} highly efficient coding scheme for neural
  networks.
\newblock {\em Parallel Processing in Neural Systems and Computers}, January
  1990.

\bibitem{butts_temporal_2007}
Daniel~A. Butts, Chong Weng, Jianzhong Jin, Chun-I. Yeh, Nicholas~A. Lesica,
  Jose-Manuel Alonso, and Garrett~B. Stanley.
\newblock Temporal precision in the neural code and the timescales of natural
  vision.
\newblock {\em Nature}, 449(7158):92--95, September 2007.
\newblock Bandiera\_abtest: a Cg\_type: Nature Research Journals Number: 7158
  Primary\_atype: Research Publisher: Nature Publishing Group.

\bibitem{gollisch_rapid_2008}
Tim Gollisch and Markus Meister.
\newblock Rapid {Neural} {Coding} in the {Retina} with {Relative} {Spike}
  {Latencies}.
\newblock {\em Science}, 319(5866):1108--1111, February 2008.
\newblock Publisher: American Association for the Advancement of Science
  Section: Report.

\bibitem{egea-weiss_high_2018}
Alexander Egea-Weiss, Alpha Renner, Christoph~J. Kleineidam, and Paul Szyszka.
\newblock High {Precision} of {Spike} {Timing} across {Olfactory} {Receptor}
  {Neurons} {Allows} {Rapid} {Odor} {Coding} in {Drosophila}.
\newblock {\em iScience}, 4:76--83, June 2018.

\bibitem{decharms_primary_1996}
R.~Christopher deCharms and Michael~M. Merzenich.
\newblock Primary cortical representation of sounds by the coordination of
  action-potential timing.
\newblock {\em Nature}, 381(6583):610--613, June 1996.
\newblock Bandiera\_abtest: a Cg\_type: Nature Research Journals Number: 6583
  Primary\_atype: Research Publisher: Nature Publishing Group.

\bibitem{bialek_reading_1991}
W.~Bialek, F.~Rieke, RR~de Ruyter~van Steveninck, and D.~Warland.
\newblock Reading a neural code.
\newblock {\em Science}, 252(5014):1854--1857, June 1991.
\newblock Publisher: American Association for the Advancement of Science
  Section: Reports.

\bibitem{panzeri_role_2001}
Stefano Panzeri, Rasmus~S. Petersen, Simon~R. Schultz, Michael Lebedev, and
  Mathew~E. Diamond.
\newblock The {Role} of {Spike} {Timing} in the {Coding} of {Stimulus}
  {Location} in {Rat} {Somatosensory} {Cortex}.
\newblock {\em Neuron}, 29(3):769--777, March 2001.
\newblock Publisher: Elsevier.

\bibitem{mackevicius_millisecond_2012}
Emily~L. Mackevicius, Matthew~D. Best, Hannes~P. Saal, and Sliman~J. Bensmaia.
\newblock Millisecond {Precision} {Spike} {Timing} {Shapes} {Tactile}
  {Perception}.
\newblock {\em Journal of Neuroscience}, 32(44):15309--15317, October 2012.

\bibitem{lawhern_spike_2011}
Vernon Lawhern, Alexandre~A. Nikonov, Wei Wu, and Robert~J. Contreras.
\newblock Spike {Rate} and {Spike} {Timing} {Contributions} to {Coding} {Taste}
  {Quality} {Information} in {Rat} {Periphery}.
\newblock {\em Frontiers in Integrative Neuroscience}, 5, 2011.
\newblock Publisher: Frontiers.

\bibitem{masquelier_learning_2013}
Timothée Masquelier and Gustavo Deco.
\newblock Learning and {Coding} in {Neural} {Networks}.
\newblock pages 513--526. May 2013.

\bibitem{masquelier_spike_2008}
Timothée Masquelier, Rudy Guyonneau, and Simon~J. Thorpe.
\newblock Spike {Timing} {Dependent} {Plasticity} {Finds} the {Start} of
  {Repeating} {Patterns} in {Continuous} {Spike} {Trains}.
\newblock {\em PLOS ONE}, 3(1):e1377, January 2008.
\newblock Publisher: Public Library of Science.

\bibitem{hayashi_modulatory_1996}
K.~Hayashi, R.~Ishikawa, L.~H. Ye, X.~L. He, K.~Takata, K.~Kohama, and
  T.~Shirao.
\newblock Modulatory role of drebrin on the cytoskeleton within dendritic
  spines in the rat cerebral cortex.
\newblock {\em The Journal of Neuroscience: The Official Journal of the Society
  for Neuroscience}, 16(22):7161--7170, November 1996.

\bibitem{wyszynski_competitive_1997}
Michael Wyszynski, Jerry Lin, Anuradha Rao, Elizabeth Nigh, Alan~H. Beggs,
  Ann~Marie Craig, and Morgan Sheng.
\newblock Competitive binding of alpha-actinin and calmodulin to the {NMDA}
  receptor.
\newblock {\em Nature}, 385(6615):439--442, January 1997.
\newblock Bandiera\_abtest: a Cg\_type: Nature Research Journals Number: 6615
  Primary\_atype: Research Publisher: Nature Publishing Group.

\bibitem{furukawa_actin-severing_1997}
K.~Furukawa, W.~Fu, Y.~Li, W.~Witke, D.~J. Kwiatkowski, and M.~P. Mattson.
\newblock The actin-severing protein gelsolin modulates calcium channel and
  {NMDA} receptor activities and vulnerability to excitotoxicity in hippocampal
  neurons.
\newblock {\em The Journal of Neuroscience: The Official Journal of the Society
  for Neuroscience}, 17(21):8178--8186, November 1997.

\bibitem{carlin_identification_1983}
R.~K. Carlin, D.~C. Bartelt, and P.~Siekevitz.
\newblock Identification of fodrin as a major calmodulin-binding protein in
  postsynaptic density preparations.
\newblock {\em The Journal of Cell Biology}, 96(2):443--448, February 1983.

\bibitem{emptage_single_1999}
N.~Emptage, T.~V. Bliss, and A.~Fine.
\newblock Single synaptic events evoke {NMDA} receptor-mediated release of
  calcium from internal stores in hippocampal dendritic spines.
\newblock {\em Neuron}, 22(1):115--124, January 1999.

\bibitem{rumelhart_parallel_1986}
David~E. Rumelhart, James~L. McClelland, and San Diego PDP Research~Group
  University~of California.
\newblock {\em Parallel distributed processing : explorations in the
  microstructure of cognition}.
\newblock Cambridge, Mass. : MIT Press, 1986.

\bibitem{hebb_organization_1949}
D.~O. Hebb.
\newblock {\em The organization of behavior; a neuropsychological theory}.
\newblock The organization of behavior; a neuropsychological theory. Wiley,
  Oxford, England, 1949.
\newblock Pages: xix, 335.

\bibitem{kohonen_self-organized_1982}
Teuvo Kohonen.
\newblock Self-organized formation of topologically correct feature maps.
\newblock {\em Biological Cybernetics}, 43(1):59--69, January 1982.

\bibitem{vaswani_attention_2017}
Ashish Vaswani, Noam Shazeer, Niki Parmar, Jakob Uszkoreit, Llion Jones,
  Aidan~N. Gomez, Lukasz Kaiser, and Illia Polosukhin.
\newblock Attention {Is} {All} {You} {Need}.
\newblock {\em arXiv:1706.03762 [cs]}, December 2017.
\newblock arXiv: 1706.03762.

\bibitem{ramsauer_hopfield_2021}
Hubert Ramsauer, Bernhard Schäfl, Johannes Lehner, Philipp Seidl, Michael
  Widrich, Thomas Adler, Lukas Gruber, Markus Holzleitner, Milena Pavlović,
  Geir~Kjetil Sandve, Victor Greiff, David Kreil, Michael Kopp, Günter
  Klambauer, Johannes Brandstetter, and Sepp Hochreiter.
\newblock Hopfield {Networks} is {All} {You} {Need}.
\newblock {\em arXiv:2008.02217 [cs, stat]}, April 2021.
\newblock arXiv: 2008.02217.

\bibitem{oesch_direction-selective_2005}
Nicholas Oesch, Thomas Euler, and W.~Rowland Taylor.
\newblock Direction-selective dendritic action potentials in rabbit retina.
\newblock {\em Neuron}, 47(5):739--750, September 2005.

\bibitem{paoletti_mechanosensitivity_1994}
P.~Paoletti and P.~Ascher.
\newblock Mechanosensitivity of {NMDA} receptors in cultured mouse central
  neurons.
\newblock {\em Neuron}, 13(3):645--655, September 1994.

\bibitem{senetar_gene_2005}
Melissa~A. Senetar and Richard~O. McCann.
\newblock Gene duplication and functional divergence during evolution of the
  cytoskeletal linker protein talin.
\newblock {\em Gene}, 362:141--152, December 2005.

\bibitem{goult_mechanical_2021}
Benjamin~T. Goult.
\newblock The {Mechanical} {Basis} of {Memory} – the {MeshCODE} {Theory}.
\newblock {\em Frontiers in Molecular Neuroscience}, 14:21, 2021.

\bibitem{martinez_complex_2003}
Luis~M. Martinez and Jose-Manuel Alonso.
\newblock {COMPLEX} {RECEPTIVE} {FIELDS} {IN} {PRIMARY} {VISUAL} {CORTEX}.
\newblock {\em The Neuroscientist : a review journal bringing neurobiology,
  neurology and psychiatry}, 9(5):317--331, October 2003.

\end{thebibliography}

\end{document}